\documentclass[final] {elsart3p}

\usepackage{amsmath}

\usepackage{placeins}

\usepackage{amssymb}

\usepackage{graphicx}

\begin{document}

\begin{frontmatter}

\title{Method of invariant grid for model reduction of hydrogen combustion}

\author{Eliodoro Chiavazzo}$^a$,
\ead{chiavazzo@lav.mavt.ethz.ch}
\author{Iliya V.\ Karlin}$^{a,b}$,
\ead{karlin@lav.mavt.ethz.ch}
\author{Christos E. Frouzakis}$^a$,
\ead{frouzakis@lav.mavt.ethz.ch}
\author{Konstantinos Boulouchos}$^a$
\ead{boulouchos@lav.mavt.ethz.ch}
\address{$^a$ Aerothermochemistry and
Combustion Systems Laboratory (LAV), ETHZ CH-8092 Zurich,
Switzerland}
\address{$^b$ School of Engineering Sciences, University of Southampton,
S017 1BJ Southampton, United Kingdom}

\begin{abstract}
The Method of Invariant Grid (MIG) is a model reduction technique based on the concept of slow invariant manifold (SIM), which approximates the SIM by a set of nodes in the concentration space (invariant grid). In the present work, the MIG is applied to a realistic combustion system: An adiabatic constant volume reactor with $H_2$-air at stoichiometric proportions. By considering the thermodynamic Lyapunov function of the detailed kinetic system, the notion of the quasi-equilibrium manifold (QEM) is adopted as an initial approximation to the SIM. One- and two-dimensional discrete approximations of the QEM (quasi-equilibrium grids) are constructed and refined via the MIG to obtain the corresponding invariant grids. The invariant grids are tabulated and used to integrate the reduced system. Excellent agreements between the reduced and detailed kinetics is demonstrated.
\end{abstract}


\end{frontmatter}

\section{Introduction}
Accurate modeling of reactive flows of hydrocarbon fuels requires the solution of a large number of conservation equations as dictated by detailed reaction mechanism. In addition to the sometimes prohibitively large number of variables introduced, the numerical solution of the governing equations has to face the stiffness introduced by the fast time scales of the kinetic terms. These issues make computations of even simple flames time consuming. On the other hand, the dynamics of complex reactive systems is often characterized by short initial transients during which the solution trajectories approach low-dimensional manifolds in the concentration space, known as the slow invariant manifolds (SIM). The remaining dynamics lasts much longer and evolves along the SIM towards the equilibrium state. The construction of the SIM allows to establish a simplified description of a complex system by extracting only the slow dynamics and neglecting the fast. As a result, the detailed large set of equations can be reduced to a much smaller system without a significant loss of accuracy. For this reason, much effort is devoted to develop automated model reduction procedures based on the concept of SIM. The methods of Intrinsic Low Dimensional Manifold (ILDM) \cite{ILDM} and Computational Singular Perturbation (CSP) \cite{CSP} represent the two examples of this family of methods.

In the sequel, the Method of Invariant Grid (MIG) introduced in \cite{1} as a computational realization of the Method of Invariant Manifold (MIM) \cite{book} is used for the first time to study a combustion problem. The Quasi-Equilibrium Grid algorithm introduced in \cite{SQEG} is adopted to obtain a first approximation of the SIM, which is afterwards refined via MIG iterations. The data delivered by that procedure are stored in tables and used to integrate a smaller and a less stiff reduced system. The results obtained with the reduced system are compared with the detailed one, and excellent agreement is found for the dynamics of all species, including the radicals, and of the temperature.

\section{Problem setup}
Consider a reactive system where $n$ chemical species $x_1,...,x_n$ participate in a complex mechanism with $r$ reactions. Let a generic reversible reaction step be written as:
\begin{equation}
\alpha _{s1} x_1 + \ldots + \alpha _{sn} x_n  \rightleftharpoons \beta _{s1} x_1  +  \ldots  + \beta _{sn} x_n ,\,
\end{equation}
where $s = 1, \ldots ,r$ is the reaction index and the integers $\alpha_{si}$ and $\beta_{si}$ are the stoichiometric coefficients of the reactants and products in the reaction $s$, respectively. Let the corresponding stoichiometric vectors be $\mbox{\boldmath$\alpha$}_s  = \left( {\alpha _{s1} ,...,\alpha _{sn} } \right)$, $\mbox{\boldmath$\beta$}_s  = \left( {\beta _{s1} ,...,\beta _{sn} } \right)$ and $\mbox{\boldmath$\gamma$}_s  = \mbox{\boldmath$\beta$}_s  - \mbox{\boldmath$\alpha$}_s$. The temporal evolution of a homogeneous reactive system is described by a system of ordinary differential equations (ODEs)
\begin{equation} \label{gen_eq}
\frac{{d\mbox{\boldmath$\psi$} }}{{dt}} = \mbox{\boldmath$g$} \left( \mbox{\boldmath$\psi$}  \right),
\end{equation}
where $\mbox{\boldmath$\psi$}$ is a set of variables that characterizes the thermo-chemical state of the system. In the sequel,
we consider an adiabatic constant volume reactor where the density and the mixture-averaged specific energy are fixed.
Let $[X_i]$ denote the molar concentration of species $i$. Once the vector $\mbox{\boldmath$\psi$} = \left( {\bar \rho ,\bar e,[X_1 ],...,[X_n ]} \right)^T$ is taken to describe any state of the system, then (\ref{gen_eq}) reads:
\begin{equation}\label{par_eq}
\frac{{d \mbox{\boldmath$\psi$} }}{{dt}} = \mbox{\boldmath$g$} \left( \mbox{\boldmath$\psi$}  \right) = \left( {0,0,\sum\limits_{s = 1}^{r} {\mbox{\boldmath$\gamma$}_s (1)\Omega _s } ,...,\sum\limits_{s = 1}^{r} {\mbox{\boldmath$\gamma$}_s (n)\Omega _s } } \right)^T,
\end{equation}
where the superscript $^T$ denotes the transposition.
The rate of reaction $s$, $\Omega_s$, and the mean internal energy can be expressed as
\begin{equation} \label{reac_rate}
\begin{array}{l}
 \Omega _s  = \Omega _s^ +   - \Omega _s^ -   \\
 \Omega _s^ +   = k_s^ +  \left( T \right)\prod\limits_{i = 1}^n {\left[ {X_i } \right]^{\alpha _i } ,\,\Omega _s^ -   = k_s^ -  \left( T \right)} \prod\limits_{i = 1}^n {\left[ {X_i } \right]^{\beta _i } }  \\
 \bar e = \sum\limits_{i = 1}^n {e_i \left( T \right)Y_i }  \\
 \end{array}
\end{equation}
where $k^+_s, k^-_s, e_i, Y_i$ are the forward and inverse reaction rate constants of the reaction $s$, and the specific internal energy and the mass fraction of species $i$, respectively. The dependence of temperature $T$ on the vector state $\mbox{\boldmath$\psi$}$ is not explicitly known and, in general, the evaluation of the right-hand side of (\ref{par_eq}) requires implicit solution of the second equation in (\ref{reac_rate}).

\subsection{Thermodynamic Lyapunov function}
Here we assume that the kinetic system (\ref{gen_eq}) describes the evolution of a chemical system towards a unique equilibrium state. Moreover, we assume that there exists a strictly convex function dependent on the state vector $\mbox{\boldmath$\psi$}$ that decreases monotonically in time under the dynamics of the system (\ref{gen_eq}). Such a function $\tilde G$ is a global Lyapunov function of the system, and it reaches the global minimum at the equilibrium point.

In a closed system under constant energy $\bar e$ and density $\bar \rho$, the value of specific mixture-averaged entropy $\bar s$ must increase monotonically starting from any initial condition. Therefore, the function $\tilde G=-\bar s$
\begin{equation}\label{entropy_Lyap}
\tilde G = \frac{{ - \sum\limits_{i = 1}^n {\left[ {s_i \left( T \right) - R\ln \left( {X_i } \right) - R\ln \left( {\frac{p}{{p_{ref} }}} \right)} \right]X_i } }}{{\bar W}},
\end{equation}
is a Lyapunov function of system (\ref{par_eq}), where $\bar W$ is the mean molecular weight, $s_i$ and $X_i$ are the entropy and the mole fraction of species $i$, respectively, $R$ is the universal gas constant while $p$ and $p_{ref}$ are the mixture total pressure and a given reference pressure, respectively. Assuming that $d$ elements are involved in the reaction, we can construct another Lyapunov function
\begin{equation}\label{entropy_Lyap_gen}
G = \tilde G + \sum\limits_{k = 1}^d {\left( {\lambda _k \sum\limits_{i = 1}^n {\mu _{ki} \left[ {X_i } \right]} } \right) + } \lambda \sum\limits_{i = 1}^n {W_i \left[ {X_i } \right]},
\end{equation}
where $\mu_{ki}$ represent the number of atoms of the $k$-th chemical element in species $i$ and $W_i$ is the molecular weight of species $i$. The fixed parameters $\lambda_k$ and $\lambda$ are chosen such that $\left. {\mbox{\boldmath$\nabla$} G} \right|_{\bar e}  = \mbox{\boldmath$0$}$ at the equilibrium, with the gradient of $G$ computed under fixed $\bar e$. Because of the conservation of atoms and density, the time derivative of (\ref{entropy_Lyap_gen}) is non-positive:
\begin{equation} \nonumber
\frac{{dG}}{{dt}} = \frac{{d\tilde G}}{{dt}} \le 0,\quad \frac{{dN_k }}{{dt}} = 0,\quad  \frac{{d\bar \rho }}{{dt}} = 0.
\end{equation}
Here, $N_k  = \sum\nolimits_{i = 1}^n {\mu_{ki} \left[ {X_i } \right]}$ denotes the total number of atoms $k$ in the system and $\bar \rho  = \sum\nolimits_{i = 1}^n {W_i \left[ {X_i } \right]}$.

\section{Theoretical background} \label{TH_Back}
In the present work, we consider the method of invariant grid (MIG) for combustion applications. In the sequel, we refer to a discrete collection of points in the concentration space as a grid. Starting from an initial approximation, the MIG approximates the slow invariant manifold (SIM) of system (\ref{par_eq}) by means of the \textit{invariant grid}, and refines the initial grid iteratively. At each iteration, the corrections are obtained by solving a system of linear equations at every node. The set of refined nodes forms a new grid, which approximates the SIM more accurately than the initial grid. Next iterations are then carried out until a termination criterion is satisfied. Finally, the reduced dynamics on the invariant grid is obtained with the help of a proper parameterization.

More details about the theoretical background of the MIG and its implementation can be found in Refs. \cite{1,book,SQEG,ChGoKa07,GKZ04}. In the sequel, we focus on the aspects related to MIG application to non-isothermal cases.

\subsection{MIG algorithm} \label{MIG_theory}
Let us consider a $q$-dimensional manifold $\mbox{\boldmath$\Omega$}$ in the $n$-dimensional concentration space. We assume that $q$ variables $\xi^i$ are associated with each point of $\mbox{\boldmath$\Omega$}$ by a smooth function $\mbox{\boldmath$F$}(\xi^1,...,\xi^q)$ which maps the variables $\xi^i$ into the concentration space. The tangent plane $\tau$ to the manifold at a given point of $\mbox{\boldmath$\Omega$}$ is spanned by $q$ vectors:
\begin{equation}\label{tangent}
\mbox{\boldmath$u$}_i  = {{\partial \mbox{\boldmath$F$}} \mathord{\left/
 {\vphantom {{\partial \mbox{\boldmath$F$}} {\partial \xi ^i }}} \right.
 \kern-\nulldelimiterspace} {\partial \xi ^i }},\quad i = 1,...,q.
\end{equation}
Any $n$-dimensional vector $\mbox{\boldmath$x$}$ can be projected onto the tangent plane $\tau$ by introducing a projector $\mbox{\boldmath$P$}$, such that the projected $n$-dimensional vector $\mbox{\boldmath$P$}(\mbox{\boldmath$x$})$ belongs to $\tau$ and $\mbox{\boldmath$P$}\left( {\mbox{\boldmath$P$}\left( \mbox{\boldmath$x$} \right)} \right) = \mbox{\boldmath$P$}\left( \mbox{\boldmath$x$} \right)$. The manifold $\mbox{\boldmath$\Omega$}$ is invariant with respect to (\ref{par_eq}) if any solution trajectory starting on $\mbox{\boldmath$\Omega$}$ proceeds towards the equilibrium state along this manifold. In other words, the difference between the vector field $\mbox{\boldmath$g$}$ and its projection on the tangent space defines the \textit{invariance condition}
\begin{equation}\label{inv_cond}
\left( {\mbox{\boldmath$I$} - \mbox{\boldmath$P$}} \right)\mbox{\boldmath$g$} = \mbox{\boldmath$0$}
\end{equation}
which must be satisfied at every point of $\mbox{\boldmath$\Omega$}$, with $\mbox{\boldmath$I$}$ denoting the identity matrix.

MIG is an iterative procedure which aims at refining an initial set of points in the concentration space (initial grid) and delivering the invariant grid that describes the slow invariant manifold. More generally, MIG regards the invariance condition (\ref{inv_cond}) as an equation and solves it by Newton-like iterations. Consider a set of nodes $\mathcal{G}$ in the concentration space (grid) approximating the $q$-dimensional SIM. Assume that at any node of $\mathcal{G}$ the $q$ tangent vectors $\mbox{\boldmath$u$}_i$ (\ref{tangent}) can be approximated by finite differences, and the construction of the projector $\mbox{\boldmath$P$}$ is defined (a specification of $\mbox{\boldmath$P$}$ will be given below). The grid $\mathcal{G}$ is considered as invariant when the left-hand side of (\ref{inv_cond}) is sufficiently small with respect to the vector field $\mbox{\boldmath$g$}$ at each node.

In order to take into account the conservation of elements and density, we consider the $(d+1) \times n$ matrix:
\begin{equation}
\mbox{\boldmath$M$} = \left[ {\begin{array}{*{20}c}
   {\mu_{11}} &  \ldots  & {\mu_{1n} }  \\
   {} &  \ldots  & {}  \\
   {\mu_{d1}} &  \ldots  & {\mu_{dn} }  \\
   {W_1 } &  \ldots  & {W_n }  \\
\end{array}} \right]
\end{equation}
Let the vector set $\left\{ {\mbox{\boldmath$b$}_1 , \ldots ,\mbox{\boldmath$b$}_h } \right\}$ be selected as a basis of the $h$-dimensional intersection of the null space of the projector $\mbox{\boldmath$P$}$ and the null space of $\mbox{\boldmath$M$}$. Where $h$ is the dimension of the latter intersection subspace. Following the MIG approach, once an initial grid $\mathcal{G}_0$ (in general, non-invariant) is given, it can be refined by solving at each node the following system of $h$ equations \cite{ChGoKa07}
\begin{equation}\label{MIG_eq}
    \sum\limits_{i = 1}^h {\delta _i \left[ {\mbox{\boldmath$b$}_j \mbox{\boldmath$J$} \mbox{\boldmath$b$}_i^T  - \mbox{\boldmath$b$}_j \mbox{\boldmath$P$} \left( {\mbox{\boldmath$J$} \mbox{\boldmath$b$}_i^T } \right)} \right] = \left[ {\mbox{\boldmath$P$} \left( \mbox{\boldmath$f$} \right) \mbox{\boldmath$b$}_j^T  - \mbox{\boldmath$f$} \mbox{\boldmath$b$}_j^T } \right]}
\end{equation}
for the unknowns $\delta_i$. The vector field $\mbox{\boldmath$f$}$ represents the time derivatives of species concentrations, $\mbox{\boldmath$f$} = \left( {{{d[X_i ]} \mathord{\left/ {\vphantom {{d[X_i ]} {dt}}} \right. \kern-\nulldelimiterspace} {dt}}} \right)$, and $\mbox{\boldmath$J$} = \left[ {{{\partial f} \mathord{\left/ {\vphantom {{\partial f} {\partial [X_i ]}}} \right. \kern-\nulldelimiterspace} {\partial [X_i ]}}} \right]$ denotes the first derivative matrix (Jacobian) of $\mbox{\boldmath$f$}$. Once the algebraic system (\ref{MIG_eq}) is solved at each point $\mbox{\boldmath$X$}^0  = \left( {[X_1^0],...,[X_n^0]} \right)$ of $\mathcal{G}_0$, a new point $\mbox{\boldmath$X$}^1$ is computed by shifting the previous one, $\mbox{\boldmath$X$}^1  = \mbox{\boldmath$X$}^0  + d \mbox{\boldmath$X$}^0$, with $d\mbox{\boldmath$X$}^0  = \sum\nolimits_{i = 1}^h {\delta _i \mbox{\boldmath$b$}_i}$, to obtain the new grid $\mathcal{G}_1$. Now, the projector can be updated on $\mathcal{G}_1$, and the MIG procedure applied for a second refinement. Iterations are terminated when, at any node, the norm of the defect of invariance $ \left| \mbox{\boldmath$\Delta$}  \right| = \left| {\mbox{\boldmath$f$} - \mbox{\boldmath$P$}\left( \mbox{\boldmath$f$} \right)} \right|$ is sufficiently small in comparison to the norm of the vector field $\left| {\mbox{\boldmath$f$}} \right|$.

Finally, it is important to discuss the projector $\mbox{\boldmath$P$}$ appearing in Eq. (\ref{MIG_eq}). The MIG method makes use of the thermodynamic projector \cite{book}, whose construction is briefly reviewed below. Let $\mbox{\boldmath$\nabla$} G$ and $\tau$ be the gradient of $G$ and the tangent hyperplane, evaluated at a given grid node $\mbox{\boldmath$X$}$, respectively. Let $\tau_0 = \tau \cap ker(\mbox{\boldmath$\nabla$} G)$, where $ker(\mbox{\boldmath$\nabla$} G)$ indicates the hyperplane orthogonal to $\mbox{\boldmath$\nabla$} G$. Assuming that $\tau  \ne \tau _0 $, let $\mbox{\boldmath$\hat u$}_1$ be a vector of the tangent plane $\tau$, such that $\mbox{\boldmath$\nabla$} G \mbox{\boldmath$\hat u$}^T_1=1$ and
\begin{equation}
\mbox{\boldmath$\hat u$}_1 \mbox{\boldmath$H$}\mbox{\boldmath$x$}^T  = 0,\quad \mbox{\boldmath$H$} = \left[ {\frac{{\partial ^2 G}}{{\partial \left[ {X_i } \right]\partial \left[ {X_j } \right]}}} \right],
\end{equation}
where $\mbox{\boldmath$x$}$ is an arbitrary vector of the subspace $\tau_0$. The thermodynamic projector acts on a generic vector $\mbox{\boldmath$\eta$}$ as follows
\begin{equation}\label{th_proj}
\mbox{\boldmath$P$} \mbox{\boldmath$\eta$}  = \left( {\mbox{\boldmath$\eta$} \mbox{\boldmath$\nabla$}G^T } \right)\mbox{\boldmath$\hat u$}_1  + \sum\nolimits_{i = 2}^n {\left( {\mbox{\boldmath$\eta$} \mbox{\boldmath$H$} \mbox{\boldmath$\hat u$}_i^T } \right) \mbox{\boldmath$\hat u$}_i }
\end{equation}
Here, the set of vectors $\left\{ {\mbox{\boldmath$\hat u$}_2 , \ldots ,\mbox{\boldmath$\hat u$}_n } \right\}$ forms a basis of $\tau_0$, such that
\begin{equation}
\mbox{\boldmath$\hat u$}_i \mbox{\boldmath$H$} \mbox{\boldmath$\hat u$}_j^T  = \delta _{ij} ,\;\forall i,j = 2,...,n
\end{equation}
with $\delta_{ij}$ denoting the Kronecker delta. In the case $\tau=\tau_0$, let $\left\{ {\mbox{\boldmath$\hat u$}_1 , \ldots ,\mbox{\boldmath$\hat u$}_n } \right\}$ be a basis of $\tau$ such that $\mbox{\boldmath$\hat u$}_i \mbox{\boldmath$H$} \mbox{\boldmath$\hat u$}_j^T  = \delta _{ij}$, then (\ref{th_proj}) takes the form:
\begin{equation}
\mbox{\boldmath$P$} \mbox{\boldmath$\eta$}  = \sum\nolimits_{i = 1}^n {\left( {\mbox{\boldmath$\eta$} \mbox{\boldmath$H$} \mbox{\boldmath$\hat u$}_i^T } \right) \mbox{\boldmath$\hat u$}_i }.
\end{equation}
It is worth noting here a remarkable feature of the thermodynamic projector: The construction of (\ref{th_proj}) on the SIM performs a slow-fast motion decomposition. In other words, close to the SIM, the slow dynamics of the system (\ref{par_eq}) takes place in the image of $\mbox{\boldmath$P$}$, while the fast dynamics evolves in its null space. More details about the Jacobian matrix $\mbox{\boldmath$J$}$, the gradient $\mbox{\boldmath$\nabla$}G$ and the second derivative matrix $\mbox{\boldmath$H$}$ are discussed in the appendix A.

\subsection{Initial approximation of the SIM}\label{QEM}
As suggested in \cite{1}, the notion of a quasi-equilibrium manifold (QEM) can be used for initializing the MIG procedure (see also \cite{ChGoKa07}). In general, a $q$-dimensional QEM represents a manifold in the concentration space which is given by minimizing the Lyapunov function (\ref{entropy_Lyap_gen}) under a set of $q$ linear constraints expressing a re-parametrization of the original variables $[X_i]$ in terms of some new variables $\xi^j$. Let us consider the minimization problem:
\begin{equation}\label{QEM_def}
\begin{array}{l}
 \min \;G \\
 s.t.\;\sum\limits_{i = 1}^n {l_i^j \left[ {X_i } \right] = \xi ^j ,\;j = 1,...,q}  \\
 \end{array}
\end{equation}
where $\mbox{\boldmath$l$}^j=(l_1^j,...,l_n^j)$ is a set of $q$ fixed vectors that will be specified below. Because of the convexity of $G$, for each fixed value of the quantities $\xi^j$, the solution of (\ref{QEM_def}) is unique if it exists \cite{Rock}. For our purpose, it proves convenient to consider the whole $q$-dimensional manifold of constrained minima, regarding the fixed quantities as parameters. Here, it is worth to point out a connection between the notion of QEM and the method of Rate Controlled Constrained Equilibrium (RCCE) \cite{RCCE71}. The RCCE method assumes that the reduced dynamics of the system (\ref{par_eq}) evolves along the QEM, obtained with a special choice of the vector set $\mbox{\boldmath$l$}^j$. Typically, one constraint concerns the total number of moles, $\sum\nolimits_{i = 1}^n {\left[ {X_i } \right] = \xi ^1}$, while others may be related, for instance, to active valence and free oxygen \cite{HamBis98}.

For our purposes, the QEM must be explicitly constructed in the concentration space and refined via the MIG procedure. Details about the special choice of the set of constraints employed here are given in the subsequent sections. The grid-based approximation of the QEM is adopted as the initial grid for the iterative procedure MIG. This approximation is constructed following the quasi-equilibrium grid algorithm (QEGA) \cite{SQEG} and proceeds as follows: Starting from an initial point $\mbox{\boldmath$X$}^0$ close to the QEM (e.g. the equilibrium itself), the function $G$ is approximated by a second-order polynomial around $\mbox{\boldmath$X$}^0$, and the minimization problem (\ref{QEM_def}) is recast to an algebraic system. The solution of this system delivers a new node $\mbox{\boldmath$X$}^1$ in the neighborhood of the former one and close to the QEM. Similarly, starting from $\mbox{\boldmath$X$}^1$, the procedure can be applied again for seeking further grid nodes, and it is terminated when negative concentrations are obtained.

\section{Application of MIG to reduction of a detailed $H_2$ mechanism}
In the sequel, a $H_2$-air system reacting according to the nine-species, $21$-step detailed mechanism of Li et al. \cite{mech04}, is studied. An adiabatic constant volume reactor with $H_2$-air mixture in stoichiometric proportions is considered, where the density and the mixture-averaged specific energy are chosen as $\bar \rho=4.58$ $kg/m^3$ and $\bar e=1.28 $ $MJ/kg$, respectively. The one- and two-dimensional slow invariant manifolds are described by constructing the pertinent invariant grids, which are utilized to integrate the reduced system. In the sequel, the concentration of species $k$ is expressed in terms of specific mole number, $\phi_k=[X_k]/ \bar \rho$.

\subsection{Thermodynamic projector}\label{TH_PR}
In the present section, we explicitly discuss the construction of the thermodynamic projector for 1-D and 2-D grids, in a nine-dimensional concentration space. Let a generic 1-D grid $\mathcal{G}$ be given as a collection of points in the concentration space. Assuming that a parameter $\xi$ is uniquely associated to any grid point, the tangent vector $\mbox{\boldmath$\hat u$} = \left( {{{d[X_1 ]} \mathord{\left/
 {\vphantom {{d[X_1 ]} {d\xi ,...,}}} \right.
 \kern-\nulldelimiterspace} {d\xi ,...,}}{{d[X_9 ]} \mathord{\left/
 {\vphantom {{d[X_9 ]} {d\xi }}} \right.
 \kern-\nulldelimiterspace} {d \xi }}} \right)$ can be approximated at any node via finite differences. Any nine-component vector $\mbox{\boldmath$\eta$}$ can be projected onto $\mathcal{G}$ as follows:
\begin{equation}\label{1dPRO}
\mbox{\boldmath$P$} \left( \mbox{\boldmath$\eta$}  \right) = \frac{1}{{\mbox{\boldmath$\nabla$} G\mbox{\boldmath$\hat u$}^T }}\left( {\mbox{\boldmath$\nabla$} G\mbox{\boldmath$\eta$} ^T } \right)\mbox{\boldmath$\hat u$}
\end{equation}
For a 2-D grid, two parameters $\xi^1$ and $\xi^2$ are associated with each point, so that two tangent vectors $\mbox{\boldmath$u$}_1 = \left( {{{\partial [X_i ]} \mathord{\left/
 {\vphantom {{\partial [X_i ]} {\partial \xi^1 }}} \right.
 \kern-\nulldelimiterspace} {\partial \xi^1 }}} \right)$ and $\mbox{\boldmath$u$}_2 = \left( {{{\partial [X_i ]} \mathord{\left/
 {\vphantom {{\partial [X_i ]} {\partial \xi^2 }}} \right.
 \kern-\nulldelimiterspace} {\partial \xi^2 }}} \right)$ can be evaluated at any grid node.
In order to construct the thermodynamic projector, it is convenient to introduce the $11 \times 12$ block matrix
\begin{equation}\label{matrix_PRO}
\mbox{\boldmath$A$} = \left[ {\begin{array}{*{20}c}
   0 & 0 & {\mbox{\boldmath$\nabla$} G}  \\
   \mbox{\boldmath$u$}_1^{T} & \mbox{\boldmath$u$}_2^{T} & {-\mbox{\boldmath$I$}}  \\
\end{array}} \right]
\end{equation}
with $\mbox{\boldmath$I$}$ denoting the identity matrix. Let us assume that $\mbox{\boldmath$A$}$ is a full rank matrix, and $\mbox{\boldmath$\hat u$}_2$ is formed by the first nine components of a vector spanning the null space of $\mbox{\boldmath$A$}$. According to the notations introduced in section \ref{MIG_theory}, let $\tau$ indicate the tangent hyperplane spanned by $\mbox{\boldmath$u$}_1$ and  $\mbox{\boldmath$u$}_2$. The intersection $\tau_0=\tau \cap ker(\mbox{\boldmath$\nabla$} G)$ is one-dimensional and $\mbox{\boldmath$\hat u$}_2$ is a basis of $\tau_0$, that is, the 2-D thermodynamic projector acts, on an arbitrary vector $\mbox{\boldmath$\eta$}$, as follows:
\begin{equation}\label{2dPRO}
\mbox{\boldmath$P$} \left( \mbox{\boldmath$\eta$}  \right) = \frac{1}{{\mbox{\boldmath$\nabla$} G \mbox{\boldmath$\hat u$}_1^T }}\left( {\mbox{\boldmath$\nabla$} G \mbox{\boldmath$\eta$}^T } \right) \mbox{\boldmath$\hat u$}_1 + \frac{1}{{\mbox{\boldmath$\hat u$}_2 \mbox{\boldmath$H$} \mbox{\boldmath$\hat u$}_2^T }}\left( {\mbox{\boldmath$\eta$} \mbox{\boldmath$H$} \mbox{\boldmath$\hat u$}_2^T } \right)\mbox{\boldmath$\hat u$}_2,
\end{equation}
where the vector $\mbox{\boldmath$\hat u$}_1$ is parallel to the hyperplane $\tau$, such that $\mbox{\boldmath$\hat u$}_1 \mbox{\boldmath$H$} \mbox{\boldmath$\hat u$}_2^T  = 0$. If $\mbox{\boldmath$A$}$ is not of full rank, we take $\mbox{\boldmath$\hat u$}_2=\mbox{\boldmath$u$}_2$ and $\mbox{\boldmath$\hat u$}_1$ parallel to $\tau$ with $\mbox{\boldmath$\hat u$}_1 \mbox{\boldmath$H$} \mbox{\boldmath$\hat u$}_2^T  = 0$.

\subsection{Construction of invariant grids}\label{MIG}

\begin{figure}
    \centering
        \includegraphics[width=0.45\textwidth]{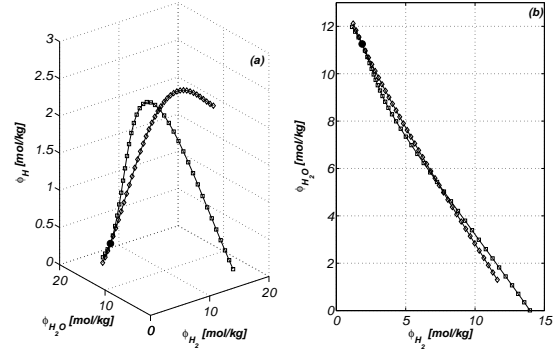}
    \caption{Starting from the equilibrium point (filled circle), the 1-D SQEG (diamonds) was constructed via QEGA and refined via MIG to obtain the 1-D invariant grid (squares).}\label{IG_SQEG_1D}
\end{figure}

\begin{figure}
    \centering
        \includegraphics[width=0.45\textwidth]{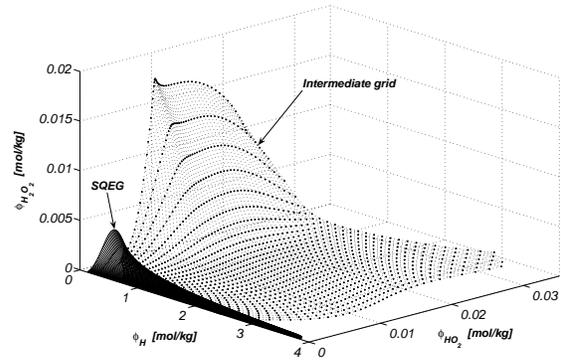}
    \caption{The 2-D SQEG (continuous lines) is constructed and refined via MIG iterations. A projection of an intermediate refined grid (dots) is shown. }\label{2D_IG_SQEG}
\end{figure}
For the case under study, a 1-D spectral quasi-equilibrium grid (SQEG) was constructed as the first approximation of the 1-D SIM. In particular, the vector $\mbox{\boldmath$l$}^1$ appearing in (\ref{QEM_def}) was taken as the left eigenvector of the Jacobian matrix $\mbox{\boldmath$J$}$ evaluated at the equilibrium point corresponding to the eigenvalue with the smallest absolute value \cite{SQEG,ChGoKa07}. The initial grid, as well as any subsequent grid, is parametrized by $\xi  = \sum\nolimits_{i = 1}^9 {l_i^1 \left[ {X_i } \right]}$. The 1-D SQEG was refined via MIG till the dimensionless ratio between the norm of the defect of invariance and the vector field ${{\left| \mbox{\boldmath$\Delta$}  \right|} \mathord{\left/ {\vphantom {{\left| \mbox{\boldmath$\Delta$}  \right|} {\left| \mbox{\boldmath$f$} \right|}}} \right. \kern-\nulldelimiterspace} {\left| \mbox{\boldmath$f$} \right|}}$ became smaller than $0.001$ at every grid node. The results are shown in Fig. \ref{IG_SQEG_1D}. Here, it is worth to mention that the SQEG and the invariant grid are in a good agreement in the neighborhood of the equilibrium point. Moreover, the SQEG also proves to be a good approximation with respect to the major species in the full concentration space (as can be seen by its projection in the $\phi_{H_2}$-$\phi_{H_2O}$ subspace, Fig. \ref{IG_SQEG_1D}b).

A 2-D SQEG was also constructed by solving the minimization problem (\ref{QEM_def}), where the two vectors $\mbox{\boldmath$l$}^1$ and $\mbox{\boldmath$l$}^2$ were chosen as the two left eigenvectors of the Jacobian matrix evaluated at the equilibrium point corresponding to the two smallest eigenvalues in absolute value. The two reduced variables associated with the grid nodes were $\xi ^1  = \sum\nolimits_{i = 1}^9 {l_i^1 \left[ {X_i } \right]}$ and $\xi ^2  = \sum\nolimits_{i = 1}^9 {l_i^2 \left[ {X_i } \right]}$. The 2-D SQEG was again refined until the threshold value $0.001$ for the ratio ${{\left| \mbox{\boldmath$\Delta$}  \right|} \mathord{\left/ {\vphantom {{\left| \mbox{\boldmath$\Delta$}  \right|} {\left| \mbox{\boldmath$f$} \right|}}} \right. \kern-\nulldelimiterspace} {\left| \mbox{\boldmath$f$} \right|}}$ was reached. If the defect of invariance at a refined node kept increasing after several iterations, the new node was discarded. The 2-D SQEG accurately describes the invariant grid only near the equilibrium point, and several Newton iterations (\ref{MIG_eq}) were required during the refinement process. Figure \ref{2D_IG_SQEG} shows both the initial SQEG grid (solid lines) and its refinement after three iterations.

The projection of the final 2-D invariant grid onto the $\phi_H$-$\phi_{HO_2}$-$\phi_{H_2O_2}$ subspace is shown in Fig. \ref{IG_SQEG_2D}. For the construction of both the 1-D and 2-D thermodynamic projector, approximation of the tangent vector $\mbox{\boldmath$u$}_j  = \left( {{{\partial \left[ {X_i } \right]} \mathord{\left/
 {\vphantom {{\partial \left[ {X_i } \right]} {\partial \xi ^j }}} \right.
 \kern-\nulldelimiterspace} {\partial \xi ^j }}} \right)$ by first-order finite differences was found to be sufficient.

\subsection{Reduced system}\label{Red_sys}
\begin{figure}
    \centering
        \includegraphics[width=0.45\textwidth]{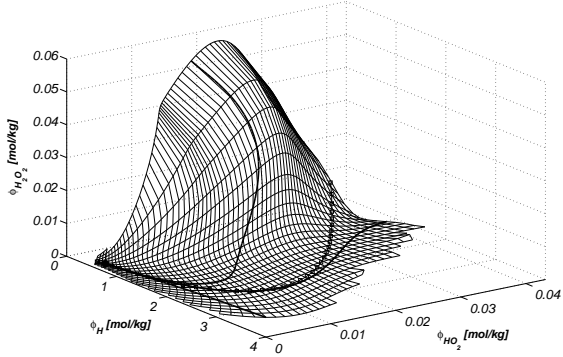}
    \caption{The 2-D invariant grid (thin lines) was computed by refining the pertinent 2-D SQEG. The 1-D invariant grid (squares) and solution trajectories (bold lines) are reported.}\label{IG_SQEG_2D}
\end{figure}
Once the invariant grid is obtained, it can be stored in tables and used during the time integration of the reduced system. Indeed,
let us assume that the 1-D invariant grid $\mathcal{G}_{inv}$ is constructed and a vector $\mbox{\boldmath$m$}=\left( {m_1 ,...,m_9 }\right)$ is chosen in such a way that the parameter $\xi=\sum\nolimits_{i = 1}^9 {m_i [X_i ]}$ is uniquely associated with every point of $\mathcal{G}_{inv}$. The original system (\ref{par_eq}) reduces to the single equation:
\begin{equation}\label{red_sys_1d}
\frac{{d \xi }}{{dt}} = \mbox{\boldmath$P$} \left( {\mbox{\boldmath$f$} \left( \xi  \right)} \right)\mbox{\boldmath$m$}^T.
\end{equation}
When a 2-D reduced description is adopted, two vectors are introduced ($\mbox{\boldmath$m$}^1$, $\mbox{\boldmath$m$}^2$) so that the new variables are $\xi^1=\sum\nolimits_{i = 1}^9 {m_i^1 [X_i ]}, \xi^2=\sum\nolimits_{i = 1}^9 {m_i^2 [X_i ]}$ and the reduced system reads:
\begin{equation}\label{red_sys_2d}
 \begin{array}{l}
 \frac{{d\xi ^1 }}{{dt}} = \mbox{\boldmath$P$} \left( {\mbox{\boldmath$f$}\left( {\xi ^1 ,\xi ^2 } \right)} \right)\mbox{\boldmath$m$}^{1T}  \\
 \frac{{d\xi ^2 }}{{dt}} = \mbox{\boldmath$P$} \left( {\mbox{\boldmath$f$}\left( {\xi ^1 ,\xi ^2 } \right)} \right)\mbox{\boldmath$m$}^{2T}  \\
 \end{array}
\end{equation}
\begin{figure}
    \centering
        \includegraphics[width=0.40\textwidth]{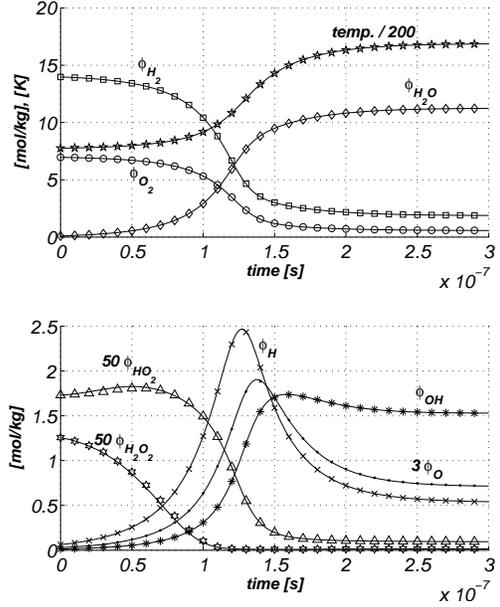}
    \caption{Starting from a point located on the 2-D invariant grid, the reduced system was integrated by using an explicit Runge-Kutta $4$-th order scheme with a fixed time step $\Delta t=10^{-8}$ s (symbols). Continuous lines represent the solution of the detailed model.}\label{evol_2D}
\end{figure}
Notice that the dependence of the vector field $\mbox{\boldmath$f$}$ on the reduced variables is not explicitly known and a proper look-up table is needed during time integration of (\ref{red_sys_1d}) or (\ref{red_sys_2d}). In other words, a continuation procedure of the approximate invariant manifold from the discrete grid has to be implemented. For the problem under study, the reduced variables of the 2-D invariant grid were chosen according to the SQEG parameterization: $\mbox{\boldmath$m$}^1=\mbox{\boldmath$l$}^1$, $\mbox{\boldmath$m$}^2=\mbox{\boldmath$l$}^2$, and the continuation of the invariant manifold from a generic $4$-node cell of the invariant grid $\mathcal{G}_{inv}$ was obtained by linear interpolation. Since the chosen parameterization of $\mathcal{G}_{inv}$ leads to a non-regular Cartesian grid in the parameter space, the $4$-node cell was mapped to a standard rectangle where a bi-variate linear interpolation was used to reconstruct the point on the SIM. The system (\ref{red_sys_2d}) was solved by an explicit $4$-th order Runge-Kutta scheme with the time step $\Delta t=10^{-8}$ s. The results were compared with the solution of the detailed system (\ref{par_eq}), obtained with the same ODE solver. However, in the latter case the time step needed was one order of magnitude smaller due to the stiffness of the detailed system. The comparison, shown in Fig. \ref{evol_2D}, proves both that (\ref{red_sys_2d}) is less stiff than (\ref{par_eq}) and a linear interpolation on the 2-D invariant grid in Fig. \ref{IG_SQEG_2D} delivers an excellent approximation of the correspondent slow invariant manifold. The results were compared on the basis of the relative deviation of the reduced solution with respect to the detailed one averaged in time. The maximum error was found to be around $2 \%$ for the evolution of $\phi_{H_2O_2}$, while for the remaining species the mean relative deviation was below $1 \%$.

Finally, it is worth to point out the computational effort needed for the construction of the invariant grid. For the case under study, the initial 2-D SQEG contains 1650 nodes. It was generated in 6 seconds and refined in about 10 minutes on a single processor 3 GHz by using a Matlab code.

\section{Conclusions}\label{conclus}
In this work, the Method of Invariant Grids (MIG) is applied for the first time to reduce a detailed hydrogen mechanism in non-isothermal conditions. The two-dimensional reduced model was then compared to the detailed one in an adiabatic constant volume reactor with $H_2$-air in stoichiometric proportions.

The Spectral Quasi Equilibrium Grid (SQEG) \cite{SQEG} proves suitable for providing the MIG with an initial collection of points (initial grid). The one- and two-dimensional SQEG grids describe quite well the dynamics of the major species, but they are not able to capture the correct evolution of some of the radicals. Therefore, several MIG iterations are needed in order to construct accurate 1-D and 2-D discrete approximation of the slow invariant manifold (SIM), providing the reduced description of the original nine-dimensional system. A bi-variate linear interpolation was used to reconstruct the SIM from the invariant grid during the time integration of the reduced system. As exemplified by the reduction in the number of time steps needed to integrate the reduced system by an order of magnitude, the stiffness is significantly reduced, and the two-dimensional models proves to be an excellent approximation of the detailed dynamics.

\section{Acknowledgments}\label{Acknow}
A. Gorban is acknowledged for the fruitful discussions and suggestions.
This work was partially supported by SNF (Project 200021-107885/1) (E.C.) and CCEM-CH (I.V.K.).

\section {Appendix A} \label{APP_MIG}
The gradient $\mbox{\boldmath$\nabla$} G$, the second derivative matrix $\mbox{\boldmath$H$}$ and the Jacobian matrix $\mbox{\boldmath$J$}$ in section \ref{MIG_theory} can be written more verbosely as follows:
\begin{equation}\label{grad_H_ex}
 \begin{array}{l}
 \mbox{\boldmath$\nabla$} G = \left( {\frac{{\partial G}}{{\partial \left[ {X_i } \right]}}} \right)_{\bar e,\left[ {X_{j \ne i} } \right]}  \\
 \mbox{\boldmath$H$} = \left[ {\frac{{\partial ^2 G}}{{\partial \left[ {X_i } \right]\partial \left[ {X_j } \right]}}} \right]_{\bar e,\left[ {X_{k \ne i,j} } \right]}  \\
 \mbox{\boldmath$J$} = \left[ {\frac{{\partial f_i }}{{\partial \left[ {X_j } \right]}}} \right]_{\bar e,\left[ {X_{k \ne j} } \right]}  \\
 \end{array}
\end{equation}
where the notation indicates that the partial derivatives are computed under fixed $\bar e$ and species concentrations. The Jacobian matrix $\mbox{\boldmath$J$}$ acts on a generic vector $\mbox{\boldmath$\eta$}$ as follows:
\begin{equation}\label{full_Jac}
\mbox{\boldmath$J$} \mbox{\boldmath$\eta$}^T  = \sum \limits_{s = 1}^{r} {\mbox{\boldmath$\gamma$}_s \left[ {\Omega _s^ +  \left( {\mbox{\boldmath$\alpha$}_s \mbox{\boldmath$H$} \mbox{\boldmath$\eta$}^T } \right) - \Omega _s^ -  \left( {\mbox{\boldmath$\beta$}_s \mbox{\boldmath$H$} \mbox{\boldmath$\eta$}^T } \right)} \right]}.
\end{equation}
At the equilibrium point, the matrix $\mbox{\boldmath$J$}$ reduces to
\begin{equation}\label{Jac_sym}
\mbox{\boldmath$J$}'(i,j) =  - \frac{1}{2}\sum\limits_{s = 1}^{r} {\left( {\Omega_s^+   + \Omega_s^-  } \right) \mbox{\boldmath$\gamma$}_s (i)} \left( {\mbox{\boldmath$H$} \mbox{\boldmath$\gamma$}_s^T } \right)(j).
\end{equation}
The latter operator is symmetric in the following sense:
\begin{equation}\label{simmetry}
\mbox{\boldmath$\eta$} \mbox{\boldmath$J$}' \mbox{\boldmath$H$} \mbox{\boldmath$\upsilon$}^T  = \mbox{\boldmath$\upsilon$} \mbox{\boldmath$J$}' \mbox{\boldmath$H$} \mbox{\boldmath$\eta$}^T,
\end{equation}
where $\mbox{\boldmath$\eta$}$ and $\mbox{\boldmath$\upsilon$}$ are two arbitrary $n$-component vectors. Because of the symmetry of $\mbox{\boldmath$J$}'$, by using the latter matrix in Eq. (\ref{MIG_eq}) at any node, instead of the full Jacobi matrix $\mbox{\boldmath$J$}$, the stability of MIG iterations can be improved \cite{GKZ04}.

Assuming that $d$ elements participate in the reaction, and $N_k$ is the total number of atoms $k$ in the system, let $\left({[Z_1],...,[Z_{n-d-1}]}\right)$ be an independent subset of $\left({[X_1],...,[X_n]} \right)$, so that the latter variables are linearly dependent on the former ones:
\begin{equation}\nonumber
    [X_i ] = [X_i ]\left( {[Z_1 ],...,[Z_{n-d-1}] , \bar \rho ,N_1,...,N_d} \right).
\end{equation}
The MIG refinements can be carried out directly in the subspace described by $[Z_i]$. The Eqs. (\ref{MIG_eq}) remain valid with the Lyapunov function in the form (\ref{entropy_Lyap}). Now, (\ref{grad_H_ex}) are substituted with
\begin{equation}
\begin{array}{l}
 \mbox{\boldmath$\nabla$} \tilde G = \left( {\frac{{\partial \tilde G}}{{\partial \left[ {Z_i } \right]}}} \right)_{\bar e,\bar \rho ,N_1 ,...,N_d ,\left[ {Z_{j \ne i} } \right]}  \\
 \mbox{\boldmath$\tilde H$} = \left[ {\frac{{\partial ^2 \tilde G}}{{\partial \left[ {Z_i } \right]\partial \left[ {Z_j } \right]}}} \right]_{\bar e,\bar \rho ,N_1 ,...,N_d ,\left[ {Z_{k \ne i,j} } \right]}  \\
 \mbox{\boldmath$J$}^*  = \left[ {\frac{{\partial f_i }}{{\partial \left[ {Z_j } \right]}}} \right]_{\bar e,\bar \rho ,N_1 ,...,N_d ,\left[ {Z_{k \ne j} } \right]}  \\
 \end{array}
\end{equation}
and the set of vectors $\{ \mbox{\boldmath$b$}_1,...,\mbox{\boldmath$b$}_h \}$, appearing in (\ref{MIG_eq}), represents a basis in the null space of the thermodynamic projector (\ref{th_proj}).
\section{Appendix B}
In this work, calculations were carried out by using the reaction mechanism suggested in \cite{mech04}, here reported in Table \ref{mechanism}.
\begin{table}[ht]
\centering
\caption{Detailed $H_2$-air reaction mechanism. Units are $cm^3-mol-sec-Kcal-K$, and $k_s(T)=A_sT^{n_s} exp(-E_s/RT)$. $^a$Troe parameter is: $F_c=0.8$. Efficiency factors are: $\varepsilon_{H_2O}=10.0$, $\varepsilon_{H_2}=1.0$ and $\varepsilon_{O_2}=-0.22$. $^b$Troe parameter is: $F_c=0.5$. Efficiency factors are: $\varepsilon_{H_2O}=11.0$, $\varepsilon_{H_2}=1.5$.} 
\begin{tabular}{l l c c c} 
\hline\hline 
Reaction &  & $A_s$ & $n_s$ & $E_s$ \\ [0.5ex] 
\hline 
1. $ H_2+O_2 \rightleftharpoons O+OH \quad $                            &       & $3.55 \times 10^{15}$ & \; -0.41 \; & \; 16.6 \; \\ 
2. $ O+H_2 \rightleftharpoons H+OH \quad $                      &           & $5.08 \times 10^{4}$  & \; 2.67 \; & \; 6.29 \; \\
3. $ H_2+OH \rightleftharpoons H_2O+H \quad $                   &           & $2.16 \times 10^{8}$  & \;1.51\; & \; 3.43 \; \\
4. $ O+H_2O \rightleftharpoons OH+OH \quad $                    &           & $2.97 \times 10^{6}$  & \;2.02\; & \;13.4\;  \\
5. $ H_2+M \rightleftharpoons H+H+M \quad $                         &           & $4.58 \times 10^{19}$ & \;-1.40\; & \;104.38\; \\
6. $ O+O+M \rightleftharpoons O_2+M \quad $                         &           & $6.16 \times 10^{15}$ & \;-0.50\; & \;0.00\; \\
7. $ O+H+M \rightleftharpoons OH+M \quad $                          &           & $4.71 \times 10^{18}$ & \;-1.0\; & \;0.00\;  \\
8. $ H+OH+M \rightleftharpoons H_2O+M \quad $                       &           & $3.8 \times 10^{22}$  & \;-2.00\; & \;0.00\; \\
9. $ H+O_2 (+M) \rightleftharpoons HO_2 (+M)^a $                  & $k_O$ & $6.37 \times 10^{20}$ & \;-1.72\; & \;0.52\; \\
10. $ HO_2+H \rightleftharpoons H_2+O_2 \quad $                     &           & $1.66 \times 10^{13}$ & \;0.00\; & \;0.82\;  \\
11. $ HO_2+H \rightleftharpoons OH+OH \quad $                           &           & $7.08 \times 10^{13}$ & \;0.00\; & \;0.30\;  \\
12. $ HO_2+O \rightleftharpoons O_2+OH \quad $                    &         & $3.25 \times 10^{13}$ & \;0.00\; & \;0.00\;  \\
13. $ HO_2+OH \rightleftharpoons H_2O+O_2 \quad $               &           & $2.89 \times 10^{13}$ & \;0.00\; & \;-0.50\; \\
14. $ HO_2+HO_2 \rightleftharpoons H_2O_2+O_2 \quad $           &           & $4.20 \times 10^{14}$ & \;0.00\; & \;11.98\; \\
15. $ HO_2+HO_2 \rightleftharpoons H_2O_2+O_2 \quad $           &           & $1.30 \times 10^{11}$ & \;0.00\; & \;-1.63\; \\
16. $ H_2O_2(+M) \rightleftharpoons 2OH(+M)^b $                     &       $k_O$ & $1.20 \times 10^{17}$ & \;0.00\; & \;45.5\;  \\
                                                                                                                    &       $k_{\infty}$ & $2.95 \times 10^{14}$ & \;0.00\;&    \;48.4\; \\
17. $ H_2O_2+H \rightleftharpoons H_2O+OH \quad $                   &           & $2.41 \times 10^{13}$ & \;0.00\; & \;3.97\;  \\
18. $ H_2O_2+H \rightleftharpoons HO_2+H_2 \quad $              &           & $4.82 \times 10^{13}$ & \;0.00\; & \;7.95\;  \\
19. $ H_2O_2+O \rightleftharpoons OH+HO_2 \quad $                   &           & $9.55 \times 10^{6}$  & \;2.00\; & \;3.97\;  \\
20. $ H_2O_2+OH \rightleftharpoons HO_2+H_2O \quad $            &           & $1.00 \times 10^{12}$ & \;0.00\; & \;0.00\;  \\
21. $ H_2O_2+OH \rightleftharpoons HO_2+H_2O \quad $            &           & $5.8 \times 10^{14}$ & \;0.00\; & \;9.56\;   \\
[1ex] 
\hline 
\end{tabular}
\label{mechanism} 
\end{table}

\begin{thebibliography}{1}
\expandafter\ifx\csname url\endcsname\relax
  \def\url#1{\texttt{#1}}\fi
\expandafter\ifx\csname urlprefix\endcsname\relax\def\urlprefix{URL }\fi
\bibitem{ILDM} U. Maas, S.B. Pope, {\em Combustion and Flames\/} 88 (1992) 239--264. 

\bibitem{CSP} S.H. Lam, D.A. Goussis, {\em International Journal of Chemical Kinetics\/} 26 (1994) 461--486.  

\bibitem{1} A.N. Gorban, I.V. Karlin, {\em Chemical Engineering Science\/} 58 (2003) 4751--4768.

\bibitem{book} A.N. Gorban, I.V. Karlin, {\em Invariant Manifolds for 
Physical and Chemical Kinetics\/}, Springer Berlin Heidelberg, 2005, p. 279.

\bibitem{SQEG} E. Chiavazzo, I.V. Karlin, {\em arXiv\/} 0704.2317 (2007). 

\bibitem{ChGoKa07} E. Chiavazzo, A.N. Gorban, I.V. Karlin, {\em Communications in Computational Physics\/} 2 (2007) 964--992. 

\bibitem{GKZ04} A.N. Gorban, I.V. Karlin, A.Y. Zinovyev, {\em Physica A\/} 333 (2004) 106--154. 

\bibitem{Rock} R.T. Rockafellar, {\em Convex Analysis\/}, Paperback edition, 1996.

\bibitem{RCCE71} J.C. Keck, D. Gillespie, {\em Combustion and Flames\/} 17 (1971) 237.

\bibitem{HamBis98} D. Hamiroune, P. Bishnu, M. Metghalchi, J.C. Keck, {\em Combustion Theory Modelling\/} 2 (1998) 81--94.

\bibitem{mech04} J. Li, Z. Zhao, A. Kazakov, F.L. Dryer, {\em International Journal of Chemical Kinetics\/} 36 (2004) 566--575.
\end{thebibliography}
\end{document}